\documentclass[aps,showpacs,nofootinbib,superscriptaddress,twocolumn]{revtex4}

\usepackage{amsmath}
\usepackage{graphicx,subfigure}
\usepackage{amssymb}
\usepackage{bm}
\usepackage{array}
\usepackage{slashed}
\usepackage{braket}
\usepackage{multirow}
\usepackage{bigstrut}
\usepackage[nooneline,footnotesize,center]{caption2}
\usepackage{longtable}
\usepackage{hyperref}

\begin{document}


\title{Nucleon strange $s\bar s$ asymmetry to the $\Lambda/\bar\Lambda$ fragmentation}


\author{Yujie Chi}
\affiliation{School of Physics and State Key Laboratory of Nuclear Physics and
Technology, Peking University, Beijing 100871,China}

\author{Xiaozhen Du}
\affiliation{School of Physics and State Key Laboratory of Nuclear Physics and
Technology, Peking University, Beijing 100871,China}

\author{Bo-Qiang Ma}
\email{mabq@pku.edu.cn}
\affiliation{School of Physics and State Key Laboratory of Nuclear Physics and
Technology, Peking University, Beijing 100871,China}
\affiliation{Collaborative Innovation Center of Quantum Matter, Beijing,China}
\affiliation{Center for High Energy Physics, Peking University, Beijing 100871,China}


\date{\today}

\begin{abstract}
The difference between the $\Lambda$ and $\bar \Lambda$ longitudinal spin transfers in the semi-inclusive deep inelastic scattering process
is intensively studied. The study is performed in the current fragmentation region, by considering the intermediate hyperon
decay processes and sea quark fragmentation processes, while the strange sea $s\bar s$ asymmetry in the nucleon is taken into
account. The calculation in the light-cone quark-diquark model
shows that the strange sea asymmetry gives a proper trend to the difference between the $\Lambda$ and
$\bar \Lambda$ longitudinal spin transfers. When considering the nonzero final hadron transverse
momentum, our results can explain the COMPASS data reasonably.
The nonzero final hadron transverse momentum is interpreted as a natural constraint to the final hadron $z$ range where
the longitudinal spin transfer is more sensitive to the strange sea $s\bar s$ asymmetry.
\end{abstract}

\pacs{14.20.Jn, 13.60.Rj, 13.87.Fh, 13.88.+e}

\maketitle

\section{Introduction}

The flavor structure of the nucleon sea, especially the existence of the strange sea $s\bar s$ asymmetry, is of great importance
in modern physics study.
In earlier studies, it is commonly assumed that the strange sea of the nucleon is particle-antiparticle $s\bar s$ symmetric, but
in fact this is manifested neither theoretically nor experimentally. In the basis of nonfundamental
symmetry violation, much progress has been made in the study of the strange sea $s\bar s$ asymmetry. We divide these studies into
two groups. One is the study mainly focusing on the nonperturbative process that
is believed to be able to produce an asymmetric intrinsic strange sea~\cite{oai:arXiv.org:hep-ph/9604393,warr:1992pr,holtmann:1996np,Ma:1997gh,Ma:1997gy,oai:arXiv.org:hep-ph/9801283,
oai:arXiv.org:hep-ph/9901321,Ma:2000uu,Alwall:2004rd,Ding:2004ht,oai:arXiv.org:hep-ph/0408292,Gao:2005gj,Wakamatsu:2004pd,oai:arXiv.org:0710.5032,Ellis:2007ig,signal:87pl,
Diehl:2007uc,oai:arXiv.org:1206.1688}.
The other is the next-to-next-to-leading order perturbative evolution process which is pointed out to be able to cause an extrinsic strange sea asymmetry~\cite{oai:arXiv.org:1206.1688,oai:arXiv.org:hep-ph/0404240}. Studies from the nonperturbative aspect have suggested that the nuclear strange sea $s\bar s$ asymmetry can give a possible explanation to the experimental CCFR data~\cite{oai:arXiv.org:hep-ex/9406007} and the NuTeV anomaly~\cite{oai:arXiv.org:hep-ex/0110059}. However, there is still no obvious evidence for the existence of the asymmetric
strange nucleon sea.

The COMPASS collaboration measured the $\Lambda$ and $\bar \Lambda$ longitudinal spin transfers in the muon-nucleon semi-inclusive deep inelastic
scattering (SIDIS) process~\cite{Alekseev:2009ab}. Their measurement shows a quite different behavior for the $\Lambda$ and $\bar \Lambda$
on the $x$ and $x_{_F}$ dependences of the longitudinal spin transfers. The spin transfer to $\Lambda$ is small, compatible to
zero, in the entire domain of the measured kinematic variables. In contrast, the longitudinal spin transfer to the $\bar \Lambda$
increases with $x_{_F}$ reaching values of $D_{LL}^{\bar \Lambda}=0.4\sim0.5$. It is pointed out in the Ref.~\cite{Ellis:2007ig} that accurate measurement of the spin transfers to the $\Lambda$ and $\bar \Lambda$ in the COMPASS kinematics has the potential to probe the intrinsic strangeness sea and their analysis mainly focuses on the target fragmentation contribution. The possibility of the strange-antistrange asymmetry contributing to the spin transfer difference of the $\Lambda$ and $\bar \Lambda$ is pointed out
in Refs.~\cite{Ma:1998pd,Ma:2000uv}, analyzed in the current fragmentation region.

In this paper, we provide a systematic study of the $\Lambda$ and $\bar \Lambda$ longitudinal spin transfer difference
in the current fragmentation region, by considering the intermediate hyperon
decay processes and sea quark fragmentation processes, while the strange sea $s\bar s$ asymmetry of the nucleon is also taken into account in a
reasonable way.

In Sec. \uppercase\expandafter{\romannumeral2}, we give the expression of the longitudinal spin transfer in a general $\ell p\rightarrow \ell P_h X$ SIDIS process. Also, we calculate the valence quark distribution functions~(PDFs) of octet baryons and the $\Sigma^{\ast}$ hyperon in the
light-cone quark-spectator-diquark model, as they are needed when we use the phenomenology Gribov-Lipatov relation to obtain the quark fragmentation functions~(FFs). Considering the CTEQ5 parametrization~\cite{Lai:1999wy} and the strange-antistrange asymmetry in the baryon-meson fluctuation model~\cite{oai:arXiv.org:hep-ph/9604393}, we present our inputs of the nucleon FFs and PDFs in the longitudinal spin transfer calculation in Sec. \uppercase\expandafter{\romannumeral3}.
In Sec. \uppercase\expandafter{\romannumeral4}, we give our results and discussions using the exact relationship of $x_{_F}$ on the $y$ and $z$ kinematical variable dependences.
We compare the calculated results with the COMPASS data. Our results indicate that the $\Lambda$ and $\bar \Lambda$ longitudinal spin transfer difference
can be explained reasonably within the light-cone $\mathrm{SU(6)}$ quark-spectator-diquark model after considering
the asymmetry between the $s$ and $\bar s$ quark distributions in the nucleon as well as the nonzero
final hadron transverse momentum contribution. Finally, we give a short summary in Sec. \uppercase\expandafter{\romannumeral5}.

\section{THE LONGITUDINAL SPIN TRANSFER AND THE LIGHT-CONE QUARK-SPECTATOR-DIQUARK MODEL }
We start from the collinear factorization theorem in the SIDIS process. From the quantum chromodynamics (QCD) factorization theorem, the high energy collision cross section can be calculated by using the perturbation theory complemented with the soft QCD effects embedded in
quark distributions and fragmentation functions, which are universal.

The differential scattering cross section at the tree level of a general $\ell p\longrightarrow \ell P_h X$ semi-inclusive deep inelastic scattering process can be expressed as~\cite{Chi:2013hka}

\begin{equation}\label{cross}
\frac{\mathrm{d}\sigma}{\mathrm{d}x\mathrm{d}y
\mathrm{d}z\mathrm{d}^2\overrightarrow{P}_{h\bot}}=
\frac{\pi\alpha_{\mathrm{em}}^2}{2Q^4}\frac{y}{z}L_{\mu\nu}W^{\mu\nu},
\end{equation}
where $x=\frac{Q^2}{2P\cdot q},~y=\frac{P\cdot q}{P\cdot \ell}$ and $z=\frac{P\cdot P_h}{P\cdot q}$ are three Lorentz invariant variables, $\overrightarrow{P}_{h\bot}$ is the transverse momentum of the produced hadron in the $\gamma^\ast P$ collinear frame. $L_{\mu\nu}$ and $W^{\mu\nu}$ are leptonic and hadronic tensors respectively and their specific forms can be referred to Ref.~\cite{Chi:2013hka}.

Then in the parton model, for a longitudinally polarized charged lepton beam and an unpolarized target, if the produced hadron is polarized, the helicity asymmetry cross section is obtained as
\begin{eqnarray}\label{asymmetry}
&&A(x,y,z)=\frac{\mathrm{d}\sigma_\Uparrow-\mathrm{d}\sigma_{\Downarrow}}
{\mathrm{d}\sigma_\Uparrow+\mathrm{d}\sigma_{\Downarrow }}\nonumber\\
&&
=\frac{\frac{4\pi\alpha_{\mathrm{em}}^2S}{Q^4}
\sum_{a}e_a^2xy(1-y/2)f_a(x,Q^2)\Delta D_a(z,Q^2)}
{\frac{4\pi\alpha_{\mathrm{em}}^2S}{Q^4}\sum_{a}e_a^2x\frac{1+(1-y)^2}{2}
f_a(x,Q^2)D_a(z,Q^2)},\nonumber\\
~
\end{eqnarray}
where the subscripts $\Uparrow$ or $\Downarrow$ denote the helicity of the produced baryon being parallel or antiparallel to the helicity of the initial incident beam, $e_a$ is the electric charge of $a$, $f_a(x,Q^2)$ is the unpolarized parton distribution function, and $D_a(z,Q^2),~\Delta D_a(z,Q^2)$ are the unpolarized and polarized fragmentation function respectively, with $a$ representing the quark or antiquark flavors, $-Q^2 = -Sxy$ being the squared 4-momentum
transfer of the virtual photon, and $S=M_p^2+m_\ell^2+2M_pE_\ell$ being the squared energy in the lepton-proton center-of-mass frame.

For a longitudinally polarized charged lepton beam and an unpolarized target, if the longitudinal polarization of the incoming lepton beam is $P_B$, the struck quark acquires a polarization $P_q = P_BD(y)$ directed along its momentum. The $D(y)$, whose explicit expression is
\begin{eqnarray}
D(y) = \frac{1-(1-y)^2}{1+(1-y)^2},
\end{eqnarray}
is the longitudinal depolarization factor taking into account the loss of polarization of the virtual photon as compared to that of the lepton. The $P_BD(y)$ distribution can be determined by subtraction of the averaged distribution of the sideband events from the distribution of the events in the signal region according to the COMPASS experiment~\cite{Alekseev:2009ab}. The spin transfer describes the probability that the polarization of the struck quark along the primary quantization axis $L$ is transferred to the $\Lambda$ hyperon along the secondary quantization axis $L^{'}$. The longitudinal spin transfer relates the produced $\Lambda$ polarization $P_L^{'}$ to the polarization of incoming lepton beam $P_B$ by ~\cite{Jaffe:1996wp}
\begin{eqnarray}
P_{L^{'}} = P_BD(y)A^{\Lambda}_{LL^{'}},
\end{eqnarray}
where $A^{\Lambda}_{LL^{'}}$ is the longitudinal spin transfer. In the COMPASS experiment, both the $L$ and $L^{'}$ are chosen along the virtual photon momentum~\cite{Alekseev:2009ab}, thus we can omit the subscripts.

In this paper, we preserve all the variables appearing in Eq.~(\ref{asymmetry}), trying to give a proper longitudinal spin transfer form. After removing the depolarization factor $D(y)$ from the asymmetry cross section, the longitudinal spin transfer is obtained as
\begin{eqnarray}\label{spintransfernew}
&&A(x,z)=\frac{\int {\rm d}y\frac{S x}{Q^4}
\sum_{a}e_a^2f_a(x,Q^2)\Delta D_a(z,Q^2)}
{\int {\rm d}y\frac{S x}{Q^4}\sum_{a}e_a^2f_a(x,Q^2)D_a(z,Q^2)}.
\end{eqnarray}
After integrating the numerator and denominator on $y$ and $x$ (or $z$) sequentially, we can obtain the longitudinal spin transfer on various kinematical variables.

When we discuss the contributions of $\Lambda$ hyperons produced from the intermediate heavier hyperon decays in SIDIS process, it is common to think that the struck
quark first fragments to various hadrons, and then some hadrons decay to $\Lambda$ according to the branching ratios that the intermediate hyperons decay to $\Lambda$. The probabilities that the struck quark fragments to various hadrons ($\Lambda,\Sigma^{0},~\Sigma^{\ast}$, etc.) are different considering the mass difference of these hadrons, and this effect should be taken into account when we calculate the contributions of the intermediate heavier hyperon decaying process according to their branching ratios. However, the probabilities that the struck quark fragments to various hadrons in the $\Lambda$ production process
are unknown to us.

In our calculation, the normalization of the fragmentation functions used in Eq.~(\ref{spintransfernew}) is chosen as $\int {\rm d}zD_a^h(z)/z=1$ for each hadron $h$, for the convenience to use the relation between the fragmentation functions and distribution functions in our discussion later. However, the real defined fragmentation functions $ D_a^h(z)$ are normalized as $\sum_{h} \int {\rm d}z zD_a^h(z)=1$. Considering this difference between these two normalizations of the fragmentation functions, there should be factors in front of the fragmentation functions we used to reflect the probabilities that the quark fragments to various hadrons. We notice that this factor in front of the total quark to $\Lambda$ fragmentation function makes no difference to our calculation since fragmentation functions appear both in the numerator and denominator in Eq.~(\ref{spintransfernew}).

The Monte Carlo calculation is used to obtain the ratios of the final $\Lambda$ hyperons produced from different channels including the direct quark fragmentation process and the intermediate hyperon decaying process. Because the probabilities that the struck quark fragments to various hadrons are already included in these ratios, it is feasible to multiplying these ratios to our calculated fragmentation functions.

A Monte Carlo calculation using the LEPTO generator
indicates that only about 40\%$-$50\% of $\Lambda$'s are produced directly,
30\%$-$40\% originate from $\Sigma^*$(1385) decay and about 20\% are
decay products of the $\Sigma^{0}$. The COMPASS collaboration
measured the relative weights of the $\Sigma^{\ast}$ and the $\Xi$ hyperon decaying to the $\Lambda \pi$.
The results are about $20\%$ smaller than those of the Monte Carlo calculation~\cite{Adolph:2013dhv}.

For the semi-inclusive $\mu p\rightarrow \mu\Lambda X$ process, when the intermediate decay process effects are considered,
the helicity-dependent fragmentation function
$\Delta D^a(z,Q^2)$ and the unpolarized fragmentation function $D^a(z,Q^2)$ of the $\Lambda$ hyperon can be reasonably written
as
\begin{eqnarray}\label{eq:po}
\Delta D_{\Lambda}^{q}(z,Q^2)&=&a_1\Delta D_{q\Lambda}(z,Q^2)+a_2\Delta D^{q}_{\Sigma^{0}}(z^\prime,Q^2)\alpha_{\Sigma^{0}\Lambda}\nonumber\\
&+&a_3\Delta D^{q}_{\Sigma^{\ast}}(z^\prime,Q^2)\alpha_{\Sigma^{\ast}\Lambda}\nonumber\\
&+&a_4\Delta D^{q}_{\Xi}(z^\prime,Q^2)\alpha_{\Xi\Lambda},
\end{eqnarray}
and
\begin{eqnarray}\label{eq:unpo}
D_{\Lambda}^{q(\bar q)}(z,Q^2)&=&a_1D_{q(\bar q)\Lambda}(z,Q^2)+a_2D^{q(\bar q)}_{\Sigma^{0}}(z^\prime,Q^2) \nonumber\\
&+&a_3D^{q(\bar q)}_{\Sigma^{\ast}}(z^\prime,Q^2)+a_4D^{q(\bar q)}_{\Xi}(z^\prime,Q^2),
\end{eqnarray}
where $\bar q$ flavors are assumed to be unpolarized in this process.

As for the $\mu p\rightarrow\mu\bar\Lambda X$, we just change the particles into their antiparticles in Eqs.~(\ref{eq:po}) and (\ref{eq:unpo}), and the same consideration should be kept in the following discussions.

The $a$'s are weight coefficients which indicate the ratios of contribution from different decay channels. Their values are adjusted as
\begin{eqnarray}\label{apa}
a_1=0.4,\quad a_2=0.2,\quad a_3=0.3,\quad a_4=0.1,
\end{eqnarray}
based on the spirit of the Monte Carlo predictions~\cite{Chi:2013hka}.

In the specific calculation, the weight coefficients of the $\Sigma^{\ast}$ hyperon are divided by three types of particles,
that are $\Sigma^{+}(1385)$, $\Sigma^{0}(1385)$ and $\Sigma^{-}(1385)$, while each type has two positively polarized spin states,
i.e., $(3/2,3/2)$ and $(3/2,1/2)$.
So the contribution to the spin transfer from the $\Sigma^{\ast}$ is
actually a mixture. To simplify
this issue, we take 10\% for each branch as an average. The same treatment is done to
the $\Xi$ hyperon, which contains the contribution from the $\Xi^{0}$ and $\Xi^{-}$, and 5\% is given to each branch.

The $\alpha$'s are decay parameters, representing the
polarization transfer from the decay hyperons to the
$\Lambda$. In our study, these parameters are set as
\begin{eqnarray}\label{alphas}
&\alpha_{\Sigma^{0}\Lambda}&=-0.333,\quad\alpha_{\Sigma^{\ast}(\frac{3}{2},\frac{3}{2})\Lambda}=1.0,\nonumber\\
&\alpha_{\Sigma^{\ast}(\frac{3}{2},\frac{1}{2})\Lambda}&=0.333,\quad\alpha_{\Xi^{0}\Lambda}=-0.406,\nonumber\\
&\alpha_{\Xi^{-}\Lambda}&=-0.458.
\end{eqnarray}

The values of $\alpha_{\Sigma^{0}\Lambda}$, $\alpha_{\Xi^{0}\Lambda}$ and $\alpha_{\Xi^{-}\Lambda}$ are taken from Refs.~\cite{Gatto:1958,Beringer:2012}, while $\alpha_{\Sigma^{\ast}\Lambda}$'s are estimated parameters by us.
The decay parameters of $\Sigma^{\ast}$ are given separately for the two types of the positive spin states. The choice of an $\alpha_{\Sigma^{\ast}(\frac{3}{2},\frac{3}{2})\Lambda}=1.0$ is
due to the facts that the spin of $\Sigma^{\ast}$ (being 3/2) should be almost
total positively correlated with $\Lambda$ spin (being 1/2)
in the decay process corresponding to the $(s,s_z)=(3/2,3/2)$ components,
and the choice of an $\alpha_{\Sigma^{\ast}(\frac{3}{2},\frac{1}{2})\Lambda}=0.333$ is the calculated result from the
$(s,s_z)=(3/2,1/2)$ components in the decay mode $\Sigma^{\ast}\rightarrow\Lambda\pi$, according to the
angular-momentum conservation law. In the decay model $\Sigma^*\rightarrow\Lambda\pi$, if the spin angular momentum of $\Sigma^{\ast}$ is
$(s,s_z)=(3/2,1/2)$, and the spin angular momentums of $\Lambda$ and $\pi$ are $(s,s_z)=(1/2,\pm 1/2)$ and $(s,s_z)=(0,0)$ respectively, the orbital angular momentum between $\Lambda$ and $\pi$ is $(L,L_z)=(1,\pm 1)$ or $(L,L_z)=(1,0)$. To obtain the $(s,s_z)=(3/2,1/2)$ component of  $\Sigma^{\ast}$, we should take
$(L,L_z)=(1,1)$ or $(L,L_z)=(1,0)$. From the Clebsch-Gordan coefficients, we know that the probability of the production of
the $(s,s_z)=(1/2, 1/2)$ component of $\Lambda$ is 2/3, and that the probability of the production of the
$(s,s_z)=(1/2, -1/2)$ component of $\Lambda$ is 1/3. Thus we get
$\alpha_{\Sigma^{\ast}(\frac{3}{2},\frac{1}{2})\Lambda} = 0.333$.

In the intermediate decay process, i.e., $q\rightarrow H_i\rightarrow \Lambda$, the longitudinal
momentum fraction of the $\Lambda$ hyperon to the splitting quark $q$ should be less than that of
the decay hyperon $H_i$ to $q$.
This can be inferred from the momentum fraction $z$ definition in the light-cone formalism,
$z=\frac{P_{\Lambda}^{-}}{q^{-}}$, for the final detected $\Lambda$ hyperon. This effect is taken into account by redefining
$\frac{P_{h}^{-}}{q^{-}}=1.1\ast\frac{P_{\Lambda}^-}{q^-}$, i.e., $z^\prime=1.1\ast z$. The relation
$z^\prime = 1.1 z$ we used is a very rough estimate based on the energy-momentum conservation law and the mass relation of the particles appearing in the intermediate hyperon decay process.

We then consider the Melosh-Wigner rotation effect in the calculation of the parton densities~\cite{Melosh:1974cu,Ma:1991xq,Ma:1992sj}, and apply the valence quark distribution functions calculated in the light-cone SU(6) quark-spectator-diquark model~\cite{Ma:1996np}
to estimate the probability of a valence quark directly
fragmenting to a hadron. This can be realized through the
phenomenology Gribov-Lipatov relation~\cite{Gribov:1971zn,Gribov:1972rt,Brodsky:1996cc,Barone:2000tx},
\begin{equation}\label{glrelation}
    D_{q}^{h}(z){\sim}zq_{h}(z),
\end{equation}
where the fragmentation function $D_{q}^{h}(z)$ indicates a quark $q$
splitting into a hadron $h$ with longitudinal momentum fraction $z$, and
the distribution function $q_{h}(z)$ presents the probability of finding
the same quark $q$ carrying longitudinal momentum fraction $x=z$ inside the same hadron $h$. Although the Gribov-Lipatov relation is only known to be valid near
the $z\rightarrow1$ and on a certain energy scale $Q^{2}_{0}$ in the leading order approximation, it is interesting to note that such a relation provides
successful descriptions of the available $\Lambda$ polarization data in several processes~\cite{Ma:2000uu,Ma:2000uv,Ma:2000cg}, based on quark distribution of the $\Lambda$ in the quark-diquark model and in the pQCD based counting rule analysis. Thus we can consider Eq.~\eqref{glrelation} as a phenomenological ansatz to parameterize the quark to $\Lambda$ fragmentation functions, and then check the validity and reasonableness of this method by
comparing the theoretical predictions with the experimental observations.

The main idea of the light-cone SU(6) quark-spectator-diquark model is to
start from the naive SU(6) wave function of the hadron and then if any one of the quarks is probed, reorganize
the other two quarks in terms of two quark wave functions with
spins 0 or 1 (scalar and vector diquarks), i.e., the diquark
serves as an effective particle which is called the spectator.

The unpolarized quark distribution for a quark with flavor $q$
inside a hadron $h$ is expressed as
\begin{equation}
q(x)= c_q^S a_S(x) + c_q^V a_V(x),
\end{equation}
where $c_q^S$ and $c_q^V$ are the weight coefficients determined by the SU(6) wave function,
and $a_D(x)$ ($D=S$ for scalar spectator or $V$ for axial vector spectator) denotes the amplitude for quark $q$ to be scattered while the
spectator is in the diquark state $D$, when expressed in terms of the
light-cone momentum space wave function $\varphi (x, {\mathbf
k}_\perp)$, reads
\begin{equation}
a_{D}(x) \propto  \int\left[\rm{d}^2 {\mathbf k}_\perp\right] |\varphi (x,
{\mathbf k}_\perp)|^2 \hspace{0.2cm}(D=S \hspace{0.1cm} \mathrm{or}
\hspace{0.1cm} V),
\end{equation}
and the normalization satisfies $\int_0^1 {\mathrm d} x a_D(x)=3$. To obtain a
practical formalism of $a_D(x)$, the
Brodsky-Huang-Lepage prescription~\cite{Brodsky:1981jv} of the
light-cone momentum space wave function for the quark-diquark is employed£º
\begin{equation}
\varphi (x, {\mathbf k}_\perp) = A_D \exp \left\{-\frac{1}{8\alpha_D^2}
\left[\frac{m_q^2+{\mathbf k}_\perp ^2}{x} + \frac{m_D^2+{\mathbf
k}_\perp^2}{1-x}\right]\right\},
\end{equation}
with the parameter $\alpha_D=330$~MeV. Other parameters in this model such as
the quark mass $m_q$, vector (scalar) diquark mass $m_{D}$
($D=S,V$) for the octet baryons are just simply estimated from the masses of the baryons.
For u and d quarks, we take $m_q\sim m_N/3$. The masses of the scalar and vector diquarks should be different taking into account the
spin force from color magnetism or alternatively from instantons~\cite{Weber:1994sq}.
QCD color-magnetic effects lift the mass degeneracy between hadrons that differ only in
the orientation of quark spins, such as $N$ and $\Delta$. The interaction is repulsive if the
spins are parallel, so that a pair of quarks in a spin-1 state (vector) has higher energy than
a pair of quarks in a spin-0 state (scalar). The energy shift between scalar and vector diquarks
produces the $N$-$\Delta$ mass splitting.  We take $m_S = 600~$MeV and $m_V = 800~$MeV
for the scalar and vector diquarks to explain the $N$-$\Delta$ mass different~\cite{Boros:1999da}.
To obtain the mass of scalar and vector diquarks containing one strange quark, we use the
phenomenological fact that the strange quark adds about 150 MeV. Thus we get $m_S = 750~$MeV
and $m_V = 950~$MeV for scalar and vector diquarks containing one strange quark; $m_S = 900~$MeV
and $m_V = 1100~$MeV for scalar and vector diquarks containing two strange quarks. The free parameters are reduced to only a few numbers, which can be referred to in Table \uppercase\expandafter{\romannumeral1}.

\vspace{0.5cm}
\begin{widetext}
\begin{center}
\begin{footnotesize}
\centerline{Table~1~~ The quark distribution functions of octet
baryons in the light-cone SU(6) quark-diquark model~\cite{Ma:2000cg}}
\vspace{0.3cm}
¡¡\renewcommand{\arraystretch}{1.3}
\begin{tabular}{cccccccc}\hline

Baryon~~~& $~~~~q ~~$ ~~& ~~~~~~~~~~~~~~~~~~~~~& $~~~~~~\Delta q ~~$
~~~~~&~~~~~~~~~~~~~~~~~~~~~~&$m_q~$(MeV) ~~~ &$m_V~$(MeV) ~~ &$m_S~$(MeV)
\\ \hline
 ~~~~p~~~~ & $~~u~~$ &$\frac{1}{6}a_V+\frac{1}{2}a_S $ &
 $\Delta u$ & $-\frac{1}{18}\tilde{a}_V+\frac{1}{2}\tilde{a}_S $ &
330 & 800 & 600 \\
 (uud)& $~~d~~$ &$\frac{1}{3}a_V$ &
 $\Delta d$ & $-\frac{1}{9}\tilde{a}_V$ &
330 & 800 & 600 \\
 ~~~~n~~~~ & $~~u~~$ &$\frac{1}{3}a_V $ &
 $\Delta u$ & $-\frac{1}{9}\tilde{a}_V $ &
330 & 800 & 600 \\
 (udd)& $~~d~~$ &$\frac{1}{6}a_V+\frac{1}{2} a_S$ &
 $\Delta d$ & $-\frac{1}{18}\tilde{a}_V+\frac{1}{2}\tilde{a}_S$ &
330 & 800 & 600 \\
 $~~~~\Sigma^{+}~~~~$ & $~~u~~$ &$\frac{1}{6}a_V+\frac{1}{2}a_S $ &
 $\Delta u$ & $-\frac{1}{18}\tilde{a}_V+\frac{1}{2}\tilde{a}_S $ &
330 & 950 & 750 \\
 (uus)& $~~s~~$ &$\frac{1}{3}a_V$ &
 $\Delta s$ & $-\frac{1}{9}\tilde{a}_V$ &
480 & 800 & 600 \\
 $~~~~\Sigma^{0}~~~~$ & $~~u~~$ &$\frac{1}{12}a_V+\frac{1}{4}a_S $ &
 $\Delta u$ & $-\frac{1}{36}\tilde{a}_V+\frac{1}{4}\tilde{a}_S $ &
330 & 950 & 750 \\
 (uds)& $~~d~~$ & $\frac{1}{12}a_V+\frac{1}{4}a_S $ &
 $\Delta d$ & $-\frac{1}{36}\tilde{a}_V+\frac{1}{4}\tilde{a}_S $ &
330 & 950 & 750 \\
 $~~~~$ & $~~s~~$ & $\frac{1}{3}a_V$ &
 $\Delta s$ & $-\frac{1}{9}\tilde{a}_V $ &
480 & 800 & 600 \\  $~~~~\Sigma^{-}~~~~$ & $~~d~~$
&$\frac{1}{6}a_V+\frac{1}{2}a_S $ &
 $\Delta d$ & $-\frac{1}{18}\tilde{a}_V+\frac{1}{2}\tilde{a}_S $ &
330 & 950 & 750 \\
 (dds)& $~~s~~$ &$\frac{1}{3}a_V$ &
 $\Delta s$ & $-\frac{1}{9}\tilde{a}_V$ &
480 & 800 & 600 \\  $~~~~\Lambda^{0}~~~~$ & $~~u~~$
&$\frac{1}{4}a_V+\frac{1}{12}a_S $ &
 $\Delta u$ & $-\frac{1}{12}\tilde{a}_V+\frac{1}{12}\tilde{a}_S $ &
330 & 950 & 750 \\
 (uds)& $~~d~~$ & $\frac{1}{4}a_V+\frac{1}{12}a_S $ &
 $\Delta d$ & $-\frac{1}{12}\tilde{a}_V+\frac{1}{12}\tilde{a}_S $ &
330 & 950 & 750 \\
 $~~~~$ & $~~s~~$ & $\frac{1}{3}a_S$ &
 $\Delta s$ & $\frac{1}{3}\tilde{a}_S $ &
480 & 800 & 600 \\
 $~~~~\Xi^{-}~~~~$ & $~~d~~$ &$\frac{1}{3}a_V $ &
 $\Delta d$ & $-\frac{1}{9}\tilde{a}_V $ &
330 & 1100 & 900 \\
 (dss)& $~~s~~$ &$\frac{1}{6}a_V+\frac{1}{2} a_S$ &
 $\Delta s$ & $-\frac{1}{18}\tilde{a}_V+\frac{1}{2}\tilde{a}_S$ &
480 & 950 & 750 \\
 $~~~~\Xi^{0}~~~~$ & $~~u~~$ &$\frac{1}{3}a_V $ &
 $\Delta u$ & $-\frac{1}{9}\tilde{a}_V $ &
330 & 1100 & 900 \\
 (uss)& $~~s~~$ &$\frac{1}{6}a_V+\frac{1}{2} a_S$ &
 $\Delta s$ & $-\frac{1}{18}\tilde{a}_V+\frac{1}{2}\tilde{a}_S$ &
480 & 950 & 750 \\ \hline
\end{tabular}
\end{footnotesize}
\end{center}
\end{widetext}

\vspace{0.5cm}

The polarized quark distributions are obtained by introducing the Melosh-Wigner correction factor
\cite{Ma:1991xq,Ma:1992sj}
\begin{equation}
\Delta q(x)= \tilde{c}_q^S \tilde{a}_S(x) + \tilde{c}_q^V
\tilde{a}_V(x),
\end{equation}
where the coefficients $\tilde{c}_q^S$ and $\tilde{c}_q^V$ are
also determined by the SU(6) quark-diquark wave function, and
$\tilde{a}_D(x)$ is expressed as
\begin{equation}
\tilde{a}_{D}(x) = \int \left[\rm{d}^2 {\mathbf k}_\perp\right]
W_D(x,{\mathbf k}_\perp) |\varphi (x, {\mathbf k}_\perp)|^2
\hspace{0.2cm} (D=S \hspace{0.1cm} or \hspace{0.1cm} V),
\end{equation}
where
\begin{equation}
W_D(x,{\mathbf k}_{\perp}) =\frac{(k^+
+m_q)^2-{\mathbf k}^2_{\perp}} {(k^+ +m_q)^2+{\mathbf
k}^2_{\perp}} \label{eqM1},
\end{equation}
with $k^+=x {\cal M}$ and ${\cal M}^2=\frac{m^2_q+{\mathbf
k}^2_{\perp}}{x}+\frac{m^2_D+{\mathbf k}^2_{\perp}}{1-x}$. The
weight coefficients are also listed in Table
\uppercase\expandafter{\romannumeral1}. In this model, though the mass difference between different quarks and diquarks breaks the SU(3) symmetry explicitly, the SU(3)
symmetry between the octet baryons is in principle maintained in formalism.

Based on the same consideration, we give the distribution functions for the
$\Sigma^{\ast}$ hyperon, which in the naive quark model is a member of the SU(3) decuplet
with the total spin of 3/2.
Here, we try to use the same parameters to estimate
both the helicity and quark distribution functions in the light-cone SU(6)
quark-spectator-diquark model based on the following reasons:
(1) the mass of $\Sigma^{\ast}$ (which is about 1385 MeV) is similar to that of
$\Xi^-$ (which is about 1321 MeV), so we can use the same effective quark mass
parameters; (2) the total quark orbital angular momentum of $\Sigma^{\ast}$ is $0$,
so to form a spin $3/2$ particle, the diquark can only be in the vector state.
The specific helicity-dependent and unpolarized quark distribution functions for
the $\Sigma^{\ast}$'s in the quark-spectator-diquark model are shown in Table \uppercase\expandafter{\romannumeral2}.

\vspace{0.5cm}
\begin{widetext}
\begin{center}
\begin{footnotesize}
\centerline{Table~2~~ The quark distribution functions of $\Sigma(1385)$'s
in the light-cone SU(6) quark-diquark model}
\vspace{0.3cm}
¡¡\renewcommand{\arraystretch}{1.5}
\begin{tabular}{cccccccc}\hline

Baryon& $~~q ~~$ ~~& ~~~~~~~~~~~~~~~& $~~\Delta q ~~$
~~&~~(3/2,3/2)~~~&~~(3/2,1/2)~~~~&$m_q~$(MeV) ~~ &$m_V~$(MeV)
\\ \hline
 $~~~~\Sigma^{+}(1385)~~~~$ & $~~u~~$ &$\frac{2}{3}a_V$ &
 $\Delta u$ & $\frac{2}{3}\tilde{a}_V$ &$\frac{2}{9}\tilde{a}_V$ &
330 & 950  \\
 (uus)& $~~s~~$ &$\frac{1}{3}a_V$ &
 $\Delta s$ & $\frac{1}{3}\tilde{a}_V$ & $\frac{1}{9}\tilde{a}_V$ &
480 & 800  \\
 $~~~~\Sigma^{0}(1385)~~~~$ & $~~u~~$ &$\frac{1}{3}a_V$ &
 $\Delta u$ & $\frac{1}{3}\tilde{a}_V$ & $\frac{1}{9}\tilde{a}_V$ &
330 & 950 \\
 (uds)& $~~d~~$ & $\frac{1}{3}a_V$ &
 $\Delta d$ & $\frac{1}{3}\tilde{a}_V$ & $\frac{1}{9}\tilde{a}_V$ &
330 & 950 \\
 $~~~~$ & $~~s~~$ & $\frac{1}{3}a_V$ &
 $\Delta s$ & $\frac{1}{3}\tilde{a}_V $ &$\frac{1}{9}\tilde{a}_V $ &
480 & 800 \\  $~~~~\Sigma^{-}(1385)~~~~$ & $~~d~~$
&$\frac{2}{3}a_V$ &
 $\Delta d$ & $\frac{2}{3}\tilde{a}_V$ &$\frac{1}{9}\tilde{a}_V$ &
330 & 950 \\
 (dds)& $~~s~~$ &$\frac{1}{3}a_V$ &
 $\Delta s$ & $\frac{1}{3}\tilde{a}_V$ &$\frac{1}{9}\tilde{a}_V$ &
480 & 800  \\ \hline

\end{tabular}
\end{footnotesize}
\end{center}
\vspace{0.5cm}
\end{widetext}
\section{THE INPUTS OF THE NUCLEON FFs AND PDFs IN THE LONGITUDINAL SPIN TRANSFER CALCULATIONS}
We know that in the naive quark model, there is an SU(3) flavor
symmetry relation between octet baryons. We consider the antiquark distribution inside
the octet baryons in the same way. To compare with the experimental data, the CTEQ5
parametrization (ctq5l) for proton~\cite{Lai:1999wy} is used as an input:
\begin{eqnarray}\label{antiquark}
u^{p}_{v}(x) &=& u^{\mathrm{ctq}}_{v}(x),\nonumber\\
d^{\Lambda}_{v}(x) &=& u^{\Lambda}_{v}(x)=\frac{u^{\Lambda,\mathrm{th}}_{v}(x)}{u^{p,\mathrm{th}}_{v}(x)}\ast u^{\mathrm{ctq}}_{v}(x),\nonumber\\
s^{\Lambda}_{v}(x) &=& \frac{s^{\Lambda,\mathrm{th}}_{v}(x)}{u^{p,\mathrm{th}}_{v}(x)}\ast u^{\mathrm{ctq}}_{v}(x),\nonumber\\
\Delta d^{\Lambda}_{v}(x) &=& \Delta u^{\Lambda}_{v}(x) = \frac{\Delta u^{\Lambda,\mathrm{th}}_{v}(x)}{u^{p,\mathrm{th}}_{v}(x)}\ast u^{\mathrm{ctq}}_{v}(x),\nonumber\\
\Delta s^{\Lambda}_{v}(x) &=& \frac{\Delta s^{\Lambda,\mathrm{th}}_{v}(x)}{u^{p,\mathrm{th}}_{v}(x)}\ast u^{\mathrm{ctq}}_{v}(x),\nonumber\\
d^{\Lambda}_{s}(x)&=&u^{\Lambda}_{s}(x) = \overline{u}^{\Lambda}(x)=\frac{1}{2}(\overline{u}^{\mathrm{ctq}}(x)+\overline{d}^{\mathrm{ctq}}(x)),\nonumber\\
s^{\Lambda}_{s}(x)&=&\overline{s}^{\Lambda}(x)=\overline{d}^{\mathrm{ctq}}(x),
\end{eqnarray}
where the $u^{\mathrm{ctq}}_{v}(x)$ means the PDF for the valence $u$ quark inside the proton from the CTEQ5L parametrization,
and the $u^{\Lambda,\mathrm{th}}_{v}(x)$ is the
PDF for the valence $u$ quark inside the $\Lambda$ given by the light-cone SU(6) quark-diquark model, so as for the other flavors.
For the other hyperons, the same spirit is followed. Applying the Gribov-Lipatov relation again, we can obtain the antiquark FFs to the same hyperon.

So far, we have given all the FFs used in the calculation. In the following, we discuss the input of the nucleon PDFs in the longitudinal spin transfer calculations.

In the baryon-meson fluctuation model~\cite{oai:arXiv.org:hep-ph/9604393}, the nucleon wave function is considered to be a
fluctuating system coupling to intermediate hadronic Fock states such as noninteracting meson-baryon pairs and the coupling
that the proton to the virtual $K^{+}\Lambda$ state
is figured out to be of most importance in the production of the intrinsic strange and antistrange asymmetric sea.
In this picture, the momentum distribution of the intrinsic $s$ and  $\bar s$ quarks can be modeled in a two-level convolution
formula:
\begin{eqnarray} \label{fluctuation}
s^{\mathrm{th}}(x)&=&\int_{x}^{1} \frac{{\rm d}y}{y}f_{\Lambda/K^+\Lambda}(y) q_{s/\Lambda}\left(\frac{x}{y}\right), \nonumber\\
\bar s^{\mathrm{th}}(x)&=&\int_{x}^{1} \frac{{\rm d}y}{y}f_{K^+/K^+\Lambda}(y) q_{\bar s/K^+}\left(\frac{x}{y}\right),
\end{eqnarray}
where $f_{\Lambda/K^+\Lambda}(y),~f_{K^+/K^+\Lambda}(y)$ are probabilities to find $\Lambda, ~K^+$ in the $K^{+}\Lambda$ state with
longitudinal momentum fraction $y$, $q_{s/\Lambda}(\frac{x}{y}),~q_{\bar s/K^+}(\frac{x}{y})$ are
probabilities to find $q,~\bar q$ in the $\Lambda,~K^{+}$ states with longitudinal momentum fraction $\frac{x}{y}$ and
these quantities can be calculated by adopting the two-body wave functions for $p=K^{+}\Lambda,~K^{+}=u\bar s, ~\Lambda=s(ud)$.
The Gaussian-type two-body wave function is
\begin{equation}\label{gaussian}
\psi_{\mathrm{Gaussian}}(\emph{M}^2)=A_{\mathrm{Gaussian}}{\rm exp}(-\frac{\emph{M}^2}{8\alpha^2}),
\end{equation}
where $\emph{M}^2=\sum_{i=1}^{2}(\mathbf{k}_{\bot i}^2+m_i^2)/x_i$ is the invariant mass of the $K^{+}\Lambda,~u\bar s ~\mathrm{or} ~s(ud)$
two-body states, and $\alpha=330~$MeV is the scaling parameter.

As is pointed out that, the fluctuation model can give the intrinsic strange sea asymmetry, which can partly explain some
important experimental phenomena, such as the strange magnetic momentum and the NuTeV anomaly etc.~\cite{Ding:2004ht,Ma:1997gh,Ma:1997gy,Ma:2000uu,
Gao:2005gj}. However, its predictions do not take into account QCD evolution effects. We also know that the CTEQ5L parametrization for the $s$ and $\bar s$ are flavor blind and the result is in fact an average. In our study, we keep the asymmetry property given by the fluctuation model while in order to reflect the evolution effects a reasonable form of the nucleon strange sea input is given as
\begin{eqnarray} \label{strangeseamodification}
s^p(x)=\frac{2s^{\mathrm{th}}(x)}{s^{\mathrm{th}(x)}+\bar s^{\mathrm{th}}(x)}s^{\mathrm{ctq}}(x), \nonumber\\
\bar s^p(x)=\frac{2\bar s^{\mathrm{th}}(x)}{s^{\mathrm{th}(x)}+\bar s^{\mathrm{th}}(x)}s^{\mathrm{ctq}}(x).
\end{eqnarray}

As for other flavors, such as $u,~d,~\bar u$ etc., the inputs are directly from the CTEQ5L parametrization.

\section{RESULTS AND DISCUSSION}
We examine the longitudinal spin transfer on the $x$ and the Feynman $x_{_F}$ variable dependences.
The calculation of the $x_{_F}$ dependent spin transfer in our formula should be done through
a kinematical transformation to relate to the $x,y ~\mathrm{and}~z$ variables.

We give the exact relationship as~(see the Appendix)
\begin{eqnarray}\label{xFrelation}
x_{_F}&=&\frac{S y z}{M\left[ M^2+S y(1-x)\right] }\left[\left(M+\frac{S y}{2M}\right)\right. \nonumber \\
&\times&\left.\sqrt{\frac{1-\left[4M^2(M_h^2+P_{h\bot}^2)\right]}{(S y z)^2}} -\sqrt{\frac{S^2 y^2}{4M^2}+S x y}\right].\nonumber\\
~
\end{eqnarray}

\begin{figure}
\includegraphics[width=1.05\columnwidth]{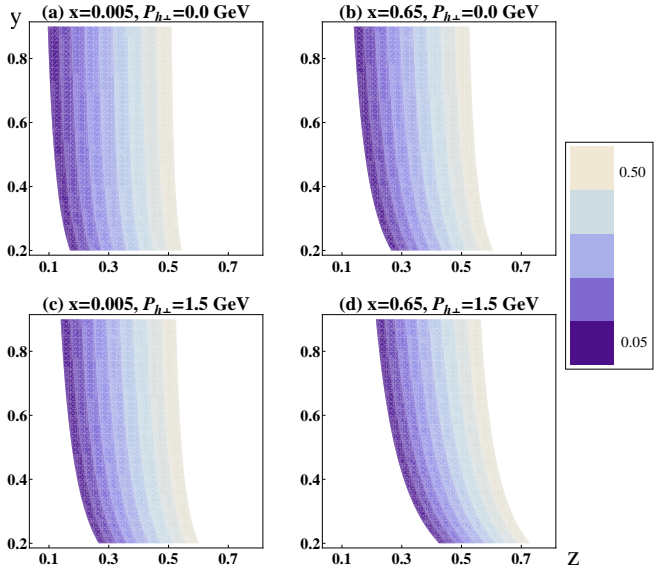}
\caption{The results of the $x_{_F}$ on the
         $y$ and $z$ kinematical variable dependences.
         The plot region of the $x_{_F}$ variable is $0.05\sim0.50$.
          The subfigures are a series of results with different
         $x$ and $P_{h\perp}$ values.}
\label{fig:xFs} 
\end{figure}

As is known, the factorization of the scattering cross section in our discussion is given in an ideal condition, that is $Q^2\rightarrow\infty,~P_{h\perp}\propto \mathcal{O}_M$. In this condition, the $P_{h\bot}$ can be neglected in Eq.~(\ref{xFrelation}).
However, the experimental data are in fact experimental condition affected, so our calculation should not be performed in an ideal way.
We try to give some nonzero value for the $P_{h\bot}$ valuable (several GeV, of order M) and find that the nonzero input of the $P_{h\bot}$
may affect the constraints between the $x_{_F},~x,~y~\mathrm{and}~z$ kinematical variables. The contour plots of these variables
are given in Fig.~\ref{fig:xFs} in the COMPASS experiment condition, where $S=320~\mathrm{GeV}^2$.
We can see from the figures that with the increase of the $P_{h\bot}$, and the increase of the $x$ variable,
at the same $x_{_F}$ numerical point, the region of the $z$ variable is significantly right shifted.

We then discuss the longitudinal spin transfer difference given by the COMPASS collaboration in two steps.
In the first step, we first set $P_{h\bot}=0$ and consider the influence from the nucleon asymmetric strange sea input, then
on the asymmetric strange sea input basis, we give two nonzero values to the $P_{h\bot}$ variable.
In the second step, we set the nucleon strange sea symmetric and see the influence from the nonzero $P_{h\bot}$ variable.
All these discussions are performed under the COMPASS experimental cuts $1~\mathrm{GeV}^2<Q^2<50~\mathrm{GeV}^2,~0.005<x<0.65,~0.2<y<0.9,~0.05<x_{_F}<0.5$ and by
the integration of equations
\begin{equation}\label{spintransferlambda}
A_{\Lambda}(x)=\frac{\int {\rm d}y {\rm d}z\frac{S x}{Q^4}
\sum_{q}e_q^2f_{q}(x,Q^2)\Delta D_{\Lambda}^{q}(z,Q^2)}
{\int {\rm d}y {\rm d}z\frac{S x}{Q^4}\sum_{q}[e_q^2f_{q}(x,Q^2)D_{\Lambda}^{q}(z,Q^2)+(q\rightarrow \bar q)]},
\end{equation}
and
\begin{equation}\label{spintransferantiLambda}
A_{\bar \Lambda}(x)=\frac{\int {\rm d}y {\rm d}z\frac{S x}{Q^4}
\sum_{\bar q}e_{\bar q}^2f_{\bar q}(x,Q^2)\Delta D_{\bar \Lambda}^{\bar q}(z,Q^2)}
{\int {\rm d}y {\rm d}z\frac{S x}{Q^4}\sum_{q}[e_q^2f_{q}(x,Q^2)D_{\bar \Lambda}^{q}(z,Q^2)+(q\rightarrow \bar q)]}.
\end{equation}

Equations~(\ref{eq:po}) to (\ref{glrelation}), (\ref{antiquark}) to (\ref{strangeseamodification}) as well as Tables \uppercase\expandafter{\romannumeral1} and \uppercase\expandafter{\romannumeral2} are all used to do the specific integrations.

\begin{figure}
\includegraphics[width=1.0\columnwidth]{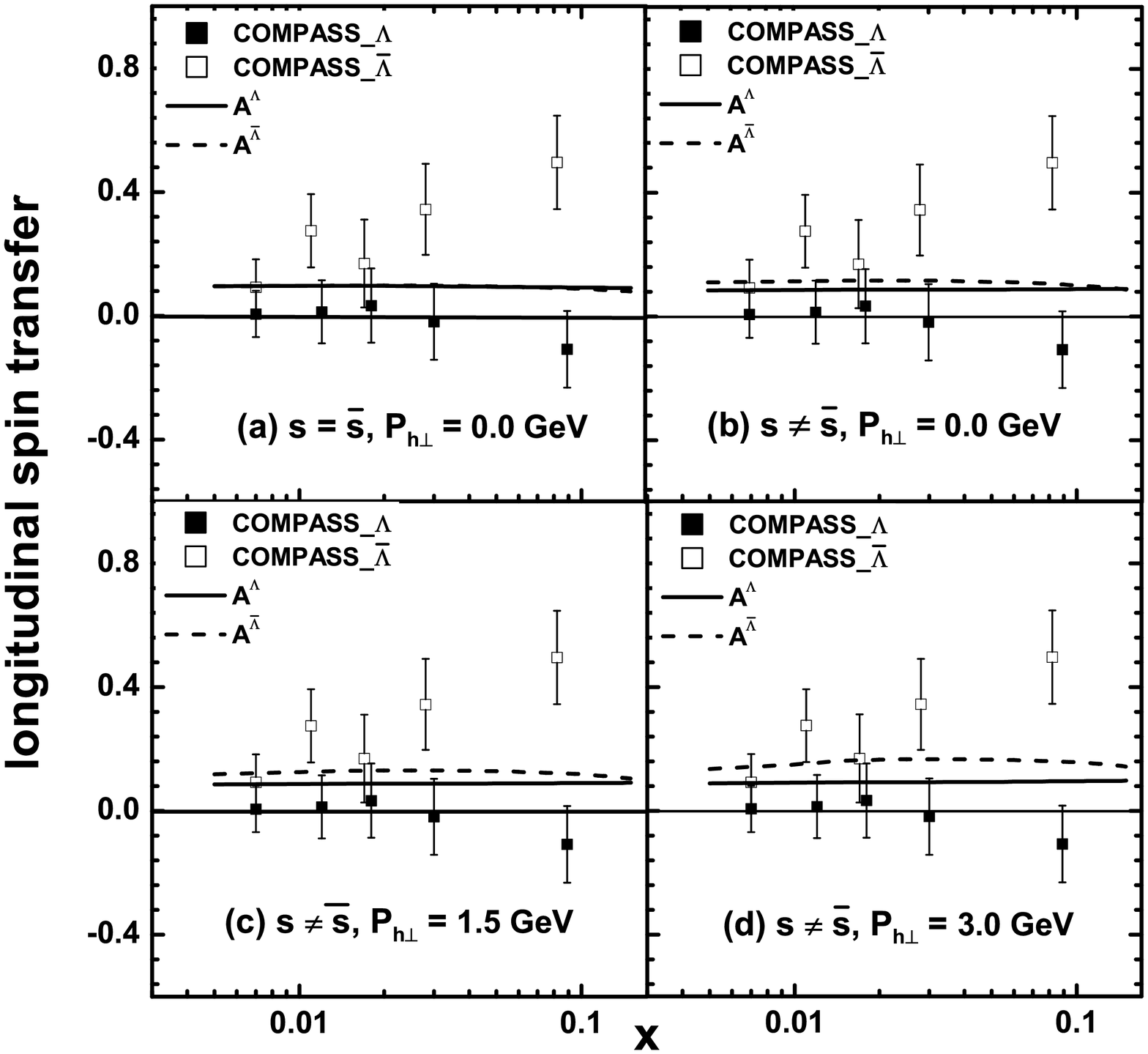}
\caption{\baselineskip 13pt The results of the
              $x$-dependent longitudinal spin transfer in the polarized charged
              lepton DIS process for the $\Lambda$ and $\bar \Lambda$ hyperons.
              Inputs of the proton strange sea asymmetry and the $P_{h\perp}$ nonzero
              values are considered step by step as shown in the subfigures.
              The data are taken from COMPASS~\cite{Alekseev:2009ab}.}
\label{fig:x} 
\end{figure}

\begin{figure}
\includegraphics[width=1.0\columnwidth]{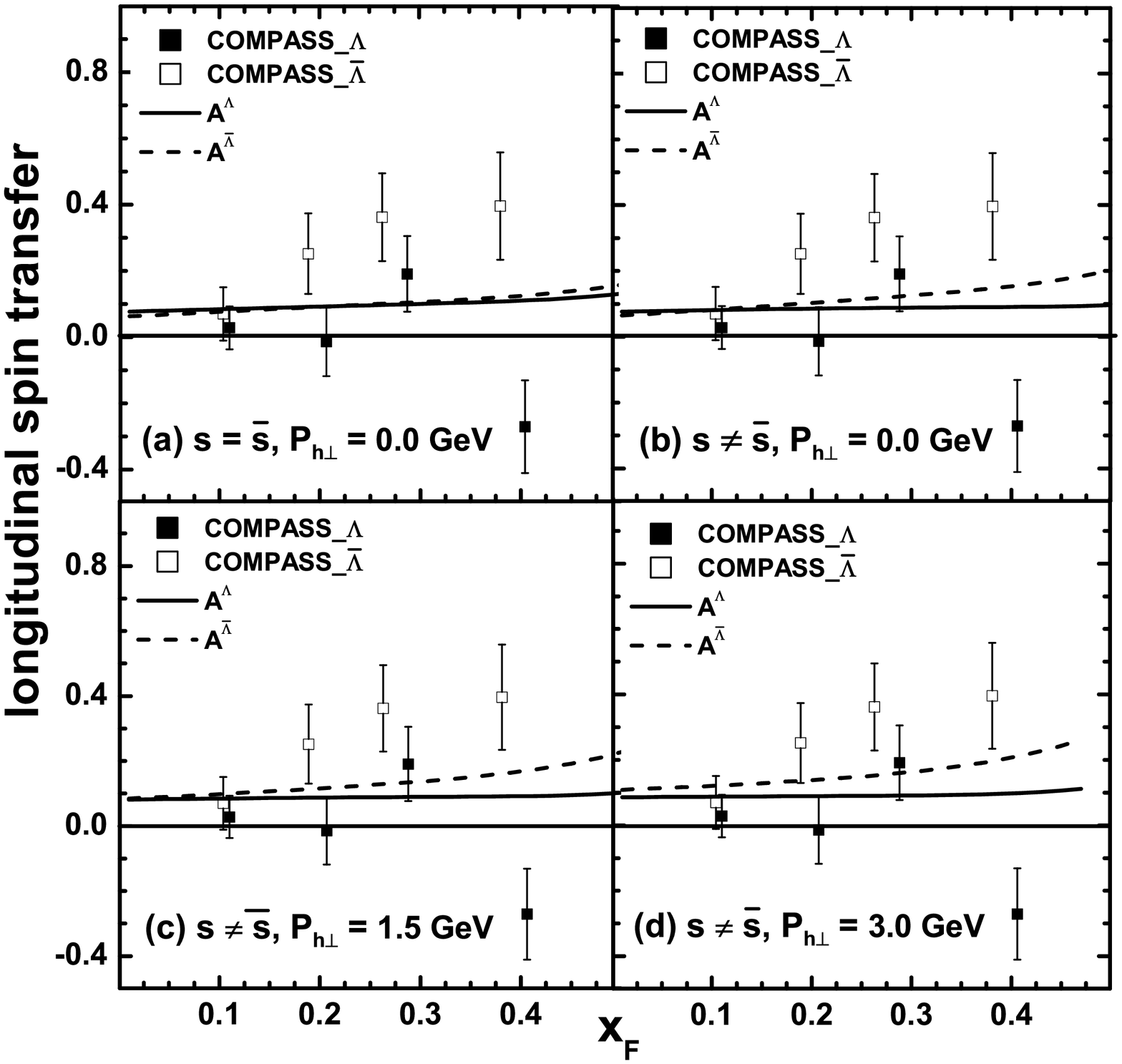}
\caption{\baselineskip 13pt The results of the
              $x_{_F}$-dependent longitudinal spin transfer in the polarized charged
              lepton DIS process for the $\Lambda$ and $\bar \Lambda$ hyperons.
              Inputs of the proton strange sea asymmetry and the $P_{h\perp}$ nonzero
              values are considered step by step as shown in the subfigures.
              The data are taken from COMPASS~\cite{Alekseev:2009ab}.}
\label{fig:xF} 
\end{figure}

\begin{figure}
\includegraphics[width=1.0\columnwidth]{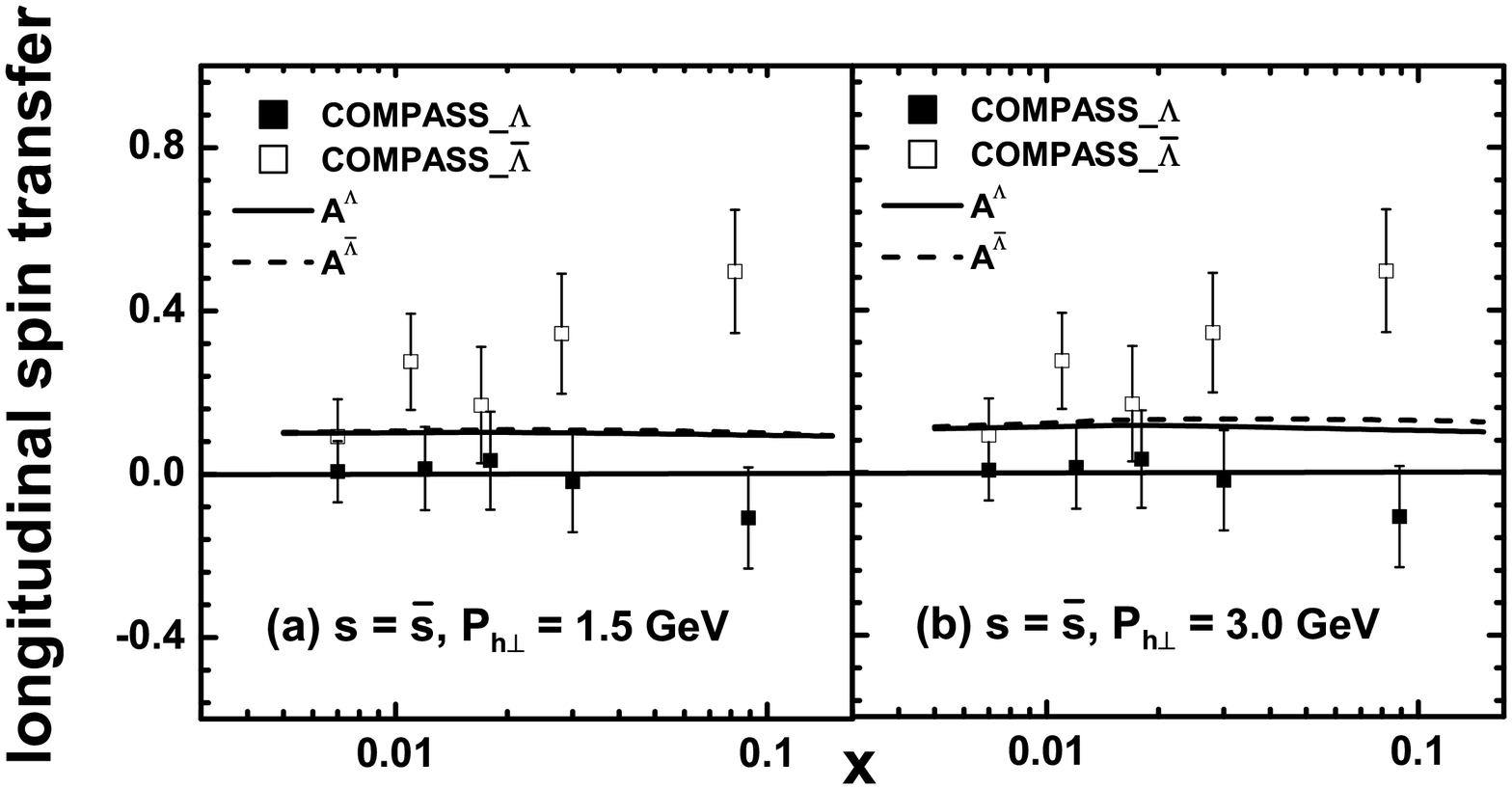}
\caption{\baselineskip 13pt The results of the
              $x$-dependent longitudinal spin transfer in the polarized charged
              lepton DIS process for the $\Lambda$ and $\bar \Lambda$ hyperons.
              The input of the proton strange sea is symmetry but the $P_{h\perp}$ is nonzero.
              The data are taken from COMPASS~\cite{Alekseev:2009ab}.}
\label{fig:xsymmetry} 
\end{figure}

\begin{figure}
\includegraphics[width=1.0\columnwidth]{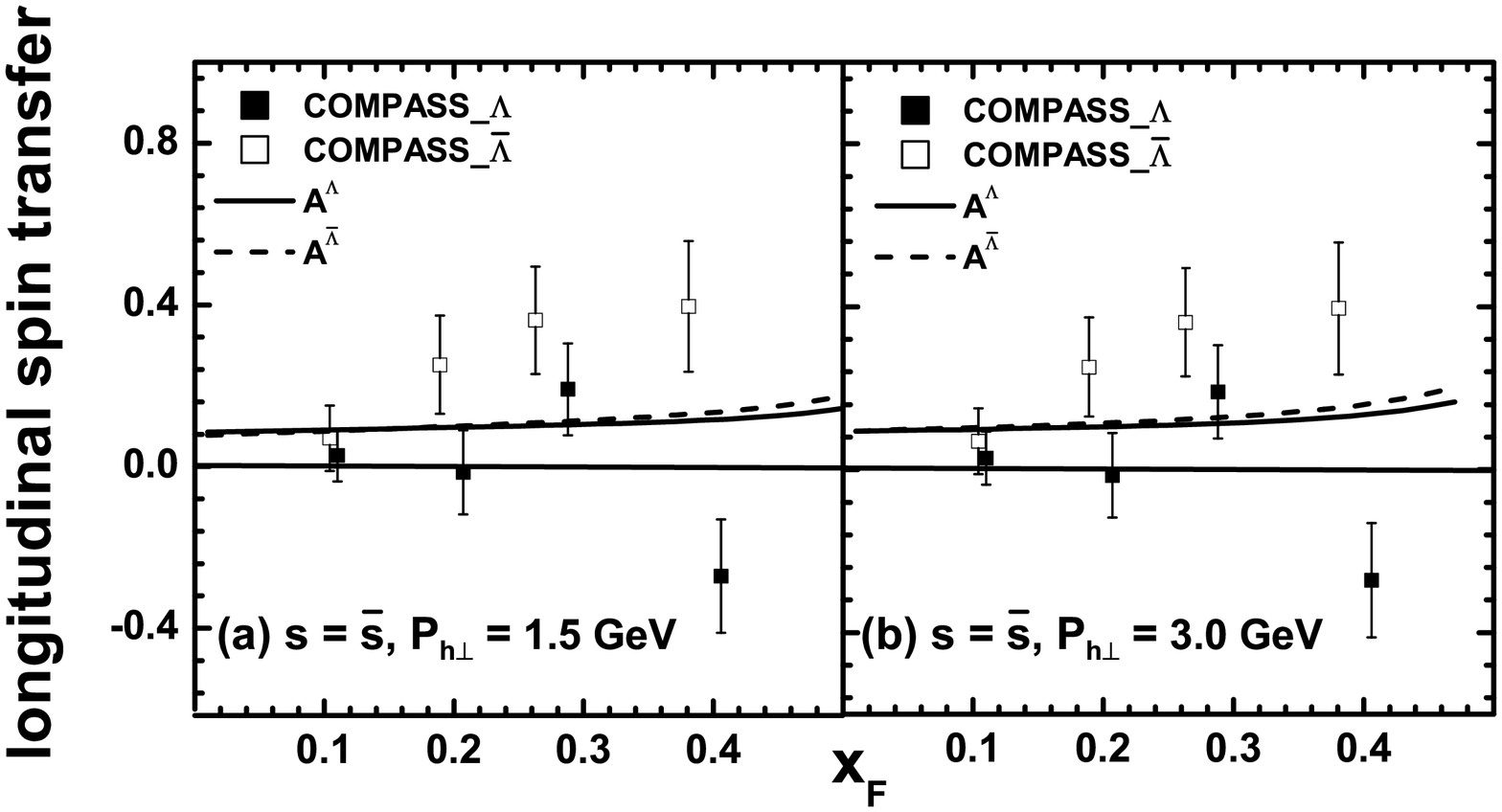}
\caption{\baselineskip 13pt The results of the
              $x_{_F}$-dependent longitudinal spin transfer in the polarized charged
              lepton DIS process for the $\Lambda$ and $\bar \Lambda$ hyperons.
              The input of the proton strange sea is symmetry but the $P_{h\perp}$ is nonzero.
              The data are taken from COMPASS~\cite{Alekseev:2009ab}.}
\label{fig:xFsymmetry} 
\end{figure}

The integration results are shown in Figs.~\ref{fig:x},~\ref{fig:xF},~\ref{fig:xsymmetry} and \ref{fig:xFsymmetry}. Among which, Figs.~\ref{fig:x} and \ref{fig:xF} are the first step calculation results of the $x$ and $x_{_F}$ variable dependences. Figure~\ref{fig:x}(a) shows the result
from the integration without considering the asymmetric sea effect or the nonzero $P_{h\perp}$ effect, while in Figs.~\ref{fig:x}(b), the asymmetric sea effect expressed in Eq.~(\ref{strangeseamodification}) is taken into account.
Figures~\ref{fig:x}(c) and \ref{fig:x}(d) are results by considering both the asymmetric sea effect and the
nonzero $P_{h\perp}$ effect. As is shown, the input of the asymmetric sea effect gives more proper trend
to the spin transfer difference than the pure integration one. Then after considering nonzero $P_{h\perp}$ values,
the difference between the spin transfers of $\Lambda$ and $\bar \Lambda$ get enlarged with an increasing
$P_{h\perp}$. At the condition of $P_{h\perp}=3.0~\mathrm{GeV}$, the result we get is qualitatively comparable with the difference of the experimental data. Compared with the experimental squared center of mass energy $S=320~\mathrm{GeV}^2$, the value of the $P_{h\perp}$ is in a reasonable region.
The similar situation appears in the $x_{_F}$-dependent spin transfer calculations, and we show the results in Figs.~\ref{fig:xF} and \ref{fig:xFsymmetry}.

Combing the first step calculation results with the $x,y ~\mathrm{and}~z$ variable constraints, we can suppose that
on the asymmetric strange sea input basis, the enlarged difference between the $\Lambda$ and $\bar \Lambda$ longitudinal
spin transfers with the increased input value of $P_{h\bot}$ is in fact from the right shifting $z$ responsible region.

Figures~\ref{fig:xsymmetry} and \ref{fig:xFsymmetry} show the second step calculation result. It is obvious that without the
asymmetric nucleon strange sea input, the influence to the spin transfer difference from the nonzero $P_{h\bot}$ is non-neglected but
still too small.

Comparing these two step discussions, we can reasonably speculate that
the large $z$ region is more sensitive to the asymmetric nucleon strange sea input, and this sensitivity can give
better explanations to the experimental data. So we suggest new and precise experimental measurement of the $\Lambda$ and $\bar \Lambda$ production in the large $z$ region to give more precise examination of the existence of the nucleon strange sea asymmetry.

\section{SUMMARY}
In summary, we studied the quark to the $\Lambda$ and $\bar \Lambda$ fragmentation properties in the current-fragmentation region
by taking various fragmentation processes into account. These processes include the intermediate decay process
and the antiquark fragmentation process, while the strange sea asymmetry in the nucleon is also taken into
account. The calculation in the light-cone quark-diquark model
shows that the strange sea $s\bar s$ asymmetry gives proper trend to the difference between the $\Lambda$ and
$\bar \Lambda$ longitudinal spin transfers. While considering the nonzero final hadron transverse
momentum, our calculation results can explain the COMPASS data reasonably.
We interpret the nonzero final hadron transverse momentum as a natural constraint to the final hadron $z$ range where
the longitudinal spin transfer is more sensitive to the strange sea asymmetry. We suggest new and precise experimental measurement of the $\Lambda$ and $\bar \Lambda$ production in the large $z$ region to make more precise examination on the nucleon strange sea
distributions.

\begin{acknowledgments}
We would like to thank Tianbo Liu and Lijing Shao for helpful discussions.
This work is partially supported by National Natural
Science Foundation of China (Grants No.~11035003, No.~11120101004 and No.~11475006) and by the Research Fund for the
Doctoral Program of Higher Education (China).
\end{acknowledgments}

\appendix
\section{}
Let us first define two Sudakov vectors $p$ and $n$ in the light-cone form as
\begin{eqnarray}
p^{\mu}=(\Lambda,0,\mathbf{0_\bot}),\quad n^{\mu}=(0,\Lambda^-,\mathbf{0_\bot}),
\end{eqnarray}
where $\Lambda$ is arbitrary.

Then the nucleon 4-momentum $P$ and the virtual photon 4-momentum $q$ in the "$\gamma^\ast N$ collinear frames"
can be represented in the form of the Sudakov vectors as
\begin{eqnarray}
P^{\mu}&=&p^{\mu}+\frac{1}{2}M^2n^{\mu}, \nonumber \\
q^{\mu}&=&\frac{Q^2}{2M^2x}\left(1-\sqrt{1+\frac{4M^2x^2}{Q^2}}\right)p^{\mu}\nonumber \\
&+&\frac{Q^2}{4x}\left(1+\sqrt{1+\frac{4M^2x^2}{Q^2}}
\right)q^{\mu},
\end{eqnarray}
where $M$ is the invariant mass of the nucleon, $x$ is the Bjoken variable and $Q^2$ is defined as $Q^2=-q^2$.

We write the 4-momentum of the final hadron $P_h$ in the $\gamma^*N$ collinear frame as
\begin{equation}
P_h^{\mu}=ap^{\mu}+bn^{\mu}+P_{h\perp}^{\mu},
\end{equation}
where $P_{h\perp}$ is the transverse vector of the final hadron which is perpendicular to the $p^{\mu}$ and $n^{\mu}$ plat.

The Lorentz invariant variable $z$ is defined as $z=P\cdot P_h/P\cdot q$ and for the final hadron it obeys
$P_h^2=M_h^2$, where $M_h$ is the final hadron invariant mass. Using these two constraints, we can get the values of $a$ and $b$.
With two variables,
\begin{eqnarray}
R&=&\sqrt{1+\frac{2M^2x}{P\cdot q}}, \nonumber \\
R^{'}&=&\sqrt{1-\frac{M^2(M_h^2+P_{h\perp}^2)}{z^2(P\cdot q)^2}},
\end{eqnarray}
$P_h$ can be written as
\begin{equation}
P_h^{\mu}=\frac{zP\cdot q}{M^2}\left(1-\frac{R^{'}}{R}\right)P^{\mu}+z\frac{R^{'}}{R}q^{\mu}+P_{h\perp}^{\mu}.
\end{equation}

Then the Feynman variable $x_{_F}$ can be obtained as
\begin{eqnarray}\label{xFrelation}
x_{_F}&=&\frac{S y z}{M\left[ M^2+S y(1-x)\right] }\left[\left(M+\frac{S y}{2M}\right)\right. \nonumber \\
&\times&\left.\sqrt{\frac{1-\left[4M^2(M_h^2+P_{h\bot}^2)\right]}{(S y z)^2}} -\sqrt{\frac{S^2 y^2}{4M^2}+S x y}\right].\nonumber\\
~
\end{eqnarray}

\end{document}